\documentclass[sigconf,authorversion,nonacm]{acmart}
\AtBeginDocument{%
  \providecommand\BibTeX{{%
    \normalfont B\kern-0.5em{\scshape i\kern-0.25em b}\kern-0.8em\TeX}}}

\setcopyright{acmlicensed}
\copyrightyear{2024}
\acmYear{2024}
\acmDOI{XXXXXXX.XXXXXXX}

\acmConference[ICPP '24]{International Conference on Parallel Processing}{August 12--15,
  2024}{Gotland, Sweden}
\acmBooktitle{Woodstock '18: ACM Symposium on Neural Gaze Detection,
 June 03--05, 2018, Woodstock, NY} 
\acmISBN{978-1-4503-XXXX-X/18/06}

\usepackage{graphicx}
\usepackage{algorithmicx}
\usepackage{textcomp}
\usepackage{xcolor}

\usepackage{multirow}
\usepackage{float}
\usepackage{listings}

\usepackage{siunitx}
\usepackage{booktabs}
\usepackage{hyperref}

\usepackage{array}

\usepackage{comment}

\usepackage{siunitx}
\usepackage{fp}
\usepackage{wrapfig}
\usepackage{subcaption}
\usepackage{sidecap}
\usepackage{graphicx}

\newcommand{\numtomillion}[1]{%
  \FPdiv{\result}{#1}{1000000}%
  \num[round-mode=places, round-precision=1, scientific-notation=false]{\result} M%
}

\newcommand{\numtothousand}[1]{%
  \FPdiv{\result}{#1}{1000}%
  \num[round-mode=places, round-precision=0, scientific-notation=false]{\result} K%
}

\newcommand{\numtothousandNoK}[1]{%
  \FPdiv{\result}{#1}{1000}%
  \num[round-mode=places, round-precision=0, scientific-notation=false]{\result} K%
}

\usepackage{tikz}

\newcommand{\poolball}[1]{%
  \begin{tikzpicture}[baseline=(current bounding box.south)]
    \filldraw[fill=black, draw=black] (0,0) circle (0.18cm);
    \filldraw[fill=white, draw=black] (0,0) circle (0.14cm);
    
    \node at (0,0) {\scriptsize\textbf{#1}};
  \end{tikzpicture}%
}

\NewDocumentCommand{\langkeyword}{m}{%
    \texttt{\textcolor{violet!80!black}{#1}}%
}

\NewDocumentCommand{\langoperator}{m}{%
    \texttt{\textcolor{orange!60!black}{#1}}%
}

\begin{document}

%\title{Structures and Techniques for Processing Highly Skewed Graphs on Message-Driven Systems}
%\title{Co-Designing A Message-Driven System for Processing Highly Skewed Graphs}
%\title{Programming Model and Runtime Support for Processing Highly Skewed Graphs on Message-Driven Systems}
\title{Structures and Techniques for Streaming Dynamic Graph Processing on Decentralized Message-Driven Systems}
\author{Bibrak Qamar Chandio}
\email{bchandio@iu.edu}
\orcid{0009-0000-8228-6273}
\affiliation{%
  \institution{\textit{Department of Intelligent Systems Engineering}\\ Indiana University Bloomington}
  %\streetaddress{P.O. Box 1212}
  %\city{Bloomington}
  \state{Indiana}
  \country{USA}
  %\postcode{47408}
}

\author{Maciej Brodowicz}
\email{mbrodowi@iu.edu}
\affiliation{%
  \institution{\textit{Department of Intelligent Systems Engineering}\\ Indiana University Bloomington}
  %\streetaddress{P.O. Box 1212}
  %\city{Bloomington}
  \state{Indiana}
  \country{USA}
  %\postcode{47408}
}

\author{Thomas Sterling}
\email{tron@iu.edu}
\affiliation{%
  \institution{\textit{Department of Intelligent Systems Engineering}\\ Indiana University Bloomington}
  %\streetaddress{P.O. Box 1212}
  %\city{Bloomington}
  \state{Indiana}
  \country{USA}
  %\postcode{47408}
}

\renewcommand{\shortauthors}{Bibrak et al.}

\begin{abstract}
The paper presents structures and techniques aimed towards co-designing scalable asynchronous and decentralized dynamic graph processing for fine-grain memory-driven architectures. It uses asynchronous active messages, in the form of \textit{actions} that send ``work to data'', with a programming and execution model that allows spawning tasks from within the data-parallelism combined with a data-structure that parallelizes vertex object across many scratchpad memory-coupled cores and yet provides a single programming abstraction to the data object.

The graph is constructed by streaming new edges using novel message delivery mechanisms and language constructs that work together to pass data and control using abstraction of actions, continuations and local control objects (LCOs) such as \textit{futures}. It results in very fine-grain updates to a hierarchical dynamic vertex data structure, which subsequently triggers a user application action to update the results of any previous computation without recomputing from scratch. In our experiments we use BFS to demonstrate our concept design, and document challenges and opportunities.
\end{abstract}


%\received{29 April 2024}
%\received[revised]{X Month 2024}
%\received[accepted]{X Month 2024}

\maketitle

\lstdefinelanguage{Racket}{
    morekeywords={struct, begin, predicate, propagate, diffuse, define, work, if, cond, let, let*, for-each, set!, cons, Integer, Float, Vector, Pointer, null, Future},
    keywordstyle={\color{violet!80!black}},
    sensitive=true,
    morecomment=[l]{;},
    morestring=[b]",
    alsoletter={<, >, !,-},
    % literate={<}{{$<$}}1 {>}{{$>$}}1 {!}{{!}}1 {lambda}{{$\lambda$}}1,
    % literate={lambda}{{$\lambda$}}1,
    morekeywords=[2]{SSSP-Action, SSSP-Diffuse, edge-addr, edge-weight, vertex-edges, vertex-level, set-vertex-level!, list, inform-neighbors, bfs-action, BFSDiffuse,  set-future!, allocate, insert-edge-action, insert-edge, vertex-has-room, vertex-ghost, vertex-score, set-vertex-score!, future-pending, enqueue-future!, future-pending!, enqueue!, rhizome-collapse, op, bcast, vertex-iteration-score, vertex-msg-count, set-vertex-msg-count!, inform-score-to-neighbors, vertex-out-degree, vertex-in-degree, not},
    keywordstyle=[2]{\color{orange!60!black}},
    morekeywords=[3]{mutable, rhizome-shared},
    keywordstyle=[3]{\color{blue}},
    literate={lambda}{{{\color{violet!80!black}$\lambda$}}}1 
              {call/cc}{{{\color{violet!80!black}call/cc}}}6 
              {eq?}{{{\color{orange!60!black}eq?}}}3 
              {?}{{{\color{orange!60!black}?}}}1 
              {'}{{{\color{red!75!black}\small\textbf{\textquotesingle}}}}1 
              {=}{{{\color{orange!60!black}=}}}1 
              {+}{{{\color{orange!60!black}+}}}1 
              %{-}{{{\color{orange!60!black}-}}}1 
              {/}{{{\color{orange!60!black}/}}}1 
              {>}{{{\color{orange!60!black}>}}}1 
              {null?}{{{\color{orange!60!black}null?}}}5 
              {\#}{{{\color{blue}\#}}}1 
              {:}{{{\color{blue}:}}}1,
}

\section{Introduction}
Streaming dynamic graph processing presents unique challenges of very fine-grain mutations to an irregular data structure representing the graph. Asynchronous and message-driven systems have the potential to naturally express these mutations in the form of active messages, that send instructions coupled with data, to where the part of the graph exists that needs to be mutated. It is in contrast to the more popular techniques of bulk synchronous models of task expression and synchronization that impose or assume a coarser-granularity of operations, and static expression of parallelism rather than dynamic discovery at runtime from the graph data itself.

\begin{figure}
  \centering
  \begin{subfigure}{0.33\linewidth}
    \centering
    \includegraphics[width=\linewidth]{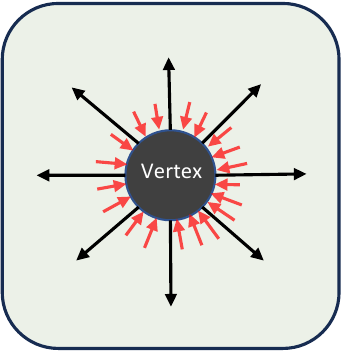}
    \caption{Logical vertex.}
    \label{fig:datastructure-vertex}
  \end{subfigure}
  \hfill
   \begin{subfigure}{0.6\linewidth}
   \centering
   \includegraphics[width=\linewidth]{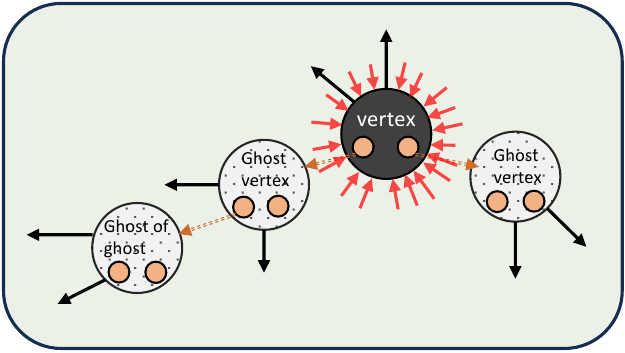}
   \caption{RPVO.}
   \label{fig:datastructure-RPVO}
  \end{subfigure}
  \hfill
  \begin{subfigure}{0.8\linewidth}
    \centering
    \includegraphics[width=\linewidth]{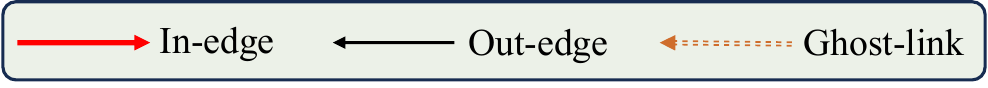}
    \captionsetup{skip=0pt} % Reduce the space between figure and caption
    \label{fig:datastructure-legend}
  \end{subfigure}
  \captionsetup{skip=0pt} % Reduce the space between figure and caption
  \caption{Vertex structures: a) The logical vertex, b) same vertex stored in a hierarchical data structure called Recursively Parallel Vertex Object (RPVO).}
    \label{fig:datastructure}
\end{figure}

\begin{figure}
  \centering
  \includegraphics[width=.8\linewidth]{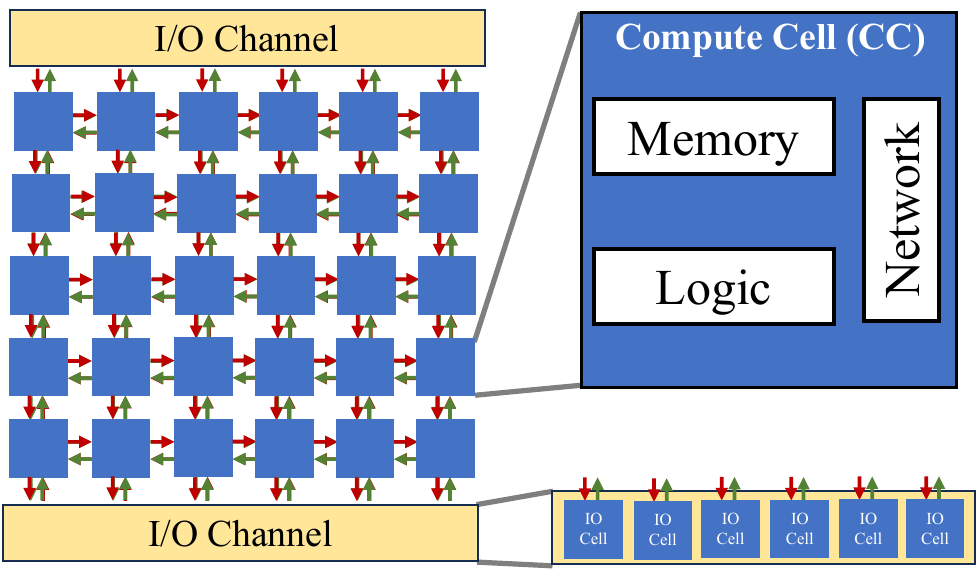}
  \captionsetup{skip=3pt} % Reduce the space between figure and caption
  \caption{A $5\times6$ AM-CCA chip shown as an exemplar. Compute Cells containing local memory along with computing logic are tessellated in a mesh network.}
\label{fig:cca-chip}
\end{figure}

This paper builds upon our previous works of a message-driven programming system \cite{chandio2024rhizomes} that parallelizes graph storage and computations, see Figure \ref{fig:datastructure}, on a highly fine-grain and asynchronous computing architecture called AM-CCA \cite{chandio2024exploring} whose high-level architectural sketch is shown in Figure \ref{fig:cca-chip}. AM-CCA is composed of homogeneous Compute Cells (CCs) having their own memory, computing ability, and neighborhood connectivity. The CCs are tessellated together, in a mesh network, to provide higher memory capacity and large amount of parallelism that is subsequently exported by a globally parallel, asynchronous, fine-grain, message-driven computing and programming model that treats the combined memory as PGAS where active messages, in the form of \textit{actions}, are sent to perform work. 

In particular, this paper focuses on streaming dynamic graph processing and contributes towards data structures and programming techniques that express and enable scalable fine-grain, asynchronous, and decentralized systems for dynamic graph processing.

\section{Message-Driven Streaming Dynamic Graphs}\label{sec:streaminggraphs}
We design and implement our message-driven streaming dynamic graph processing using the ``\textit{diffusive programming model}'', explained in \cite{chandio2024rhizomes}, under which an asynchronous active message, in the form of an \textit{action}, is sent from a memory locality to another memory locality (target). The memory locality can either be on the same CC or on a different CC. This \textit{action} can mutate the state of the target locality and can further create new \textit{actions} (work) at the destination thereby creating a ripple effect or \textit{diffusion}.

\lstset{
  language=C++,
  basicstyle=\ttfamily\footnotesize,
  keywordstyle=\color{blue}\ttfamily,
  stringstyle=\color{red}\ttfamily,
  commentstyle=\color{green!40!black}\ttfamily,
  morecomment=[l][\color{magenta}]{\#},
  showstringspaces=false,
  breaklines=true,
  %backgroundcolor=\color{gray!10},    % Background color
  numbers=left,
  numberstyle=\tiny\color{gray},
  numbersep=2pt,
  upquote=true,
}

\begin{lstlisting}[language=C++,caption=Pseudocode for a typical main() that orchestrates the data transfer to the device.,label=lst:main,xleftmargin=0.3cm]
void main() {
  AMCCA_Device dev = /* Initialize the device. */
    
  std::map<int, Pointer> vertices = /* allocate
               vertices on the device and 
               get their addresses. */

  std::vector<Edge> edges = /* get edges. */

  // Register the `insert-edge-action` action.
  AMCCA_REGISTER_ACTION(dev, 
                        INSERT_ACTION, 
                        "insert-edge-action");

  // Register the edge transfer with IO channels.
  dev.register_data_transfer(vertices, 
                             edges,
                             INSERT_ACTION);

  // Create a termintor object that handles 
  // termination detection for the diffusion.
  AMCCA_Terminator terminator = AMCCA_Terminator();                                       

  // Diffuse and wait on the terminator.
  dev.run(terminator);
}
\end{lstlisting}

Our streaming dynamic graph edge ingestion is implemented in the \langoperator{insert-edge-action} \textit{action} of Listing \ref{lst:cca-insert-edge-simple}. The first operand, \texttt{v}, is the memory address of the vertex on which this \textit{action} is invoked. Details of the vertex type are shown in Listing \ref{lst:data-structure-edge} and Listing \ref{lst:data-structure-vertex}. To enable dynamic streaming BFS, when an edge is inserted in a vertex, it passes the BFS level along using \langoperator{bfs-action} \textit{action} of Listing \ref{lst:cca-bfs}.

Listing \ref{lst:main} shows a typical call to an AM-CCA diffusive program in a manner of an accelerator. The edges are read by the IO channels, which then distribute them among their respective IO Cells. When the computation starts, every cycle, each IO Cell reads an edge, creates the corresponding action registered with \texttt{INSERT\_ACTION}, and sends it to its connected CC.

\begin{figure}
  \begin{minipage}[l]{0.3\textwidth}
\begin{lstlisting}[language=Racket,caption=Vertex-Centric Data Structure,label=lst:data-structure-vertex,xleftmargin=0.3cm]
; vertex type for BFS
(struct vertex 
  ([id     : Integer] 
   [level  : Integer]
   [edges  : (Vector edge)]
   [ghosts : (Vector 
              (Future Pointer))]))
\end{lstlisting}
    %\label{fig:cca-chip-memory-controller}
  \end{minipage}\hfill
  \begin{minipage}[l]{0.17\textwidth}
\begin{lstlisting}[language=Racket,caption=Edge type,label=lst:data-structure-edge,xleftmargin=0.2cm]
; edge type
(struct edge 
  ([addr : Pointer]
   [w  : Integer]))
\end{lstlisting}
  \end{minipage}
\end{figure}

\section{Synchronization}
The computing model is based on global parallelism using a decentralized, event-driven, and asynchronous execution. Data synchronization and conditional control transfer is achieved using Local Control Objects (LCOs), of ParalleX \cite{paralleX2009}\cite{paralleX2007} and HPX \cite{SurveyAMT2017}. The use of LCOs preserves global parallelism and fine-granularity. In this paper, we use the \textit{future} LCO to implement data ingestion. It is used when the local edge-list of a vertex (or ghost vertex) is full and a new ghost vertex must be allocated before inserting the edge. The ghost vertex pointer, that has the type \textit{future} of \langkeyword{Pointer} type, is set by a continuation that returns with the address of newly allocated memory. Section \ref{subsec:continuation-future} provides design details of our approach.

\begin{figure*}
  \begin{minipage}[t]{0.11\textwidth}
    \includegraphics[width=\textwidth]{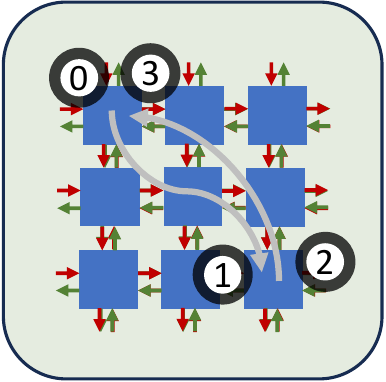}
  \end{minipage}\hfill
  \begin{minipage}[b]{0.87\textwidth}
    \caption{Asynchronous control transfer. \protect\poolball{0} Runtime sends a system action \langoperator{allocate}, configured with a return trigger action, to a remote compute cell. \protect\poolball{1} the remote compute cell allocates memory. \protect\poolball{2} memory address is sent back in the form of the trigger action that is targeted originating vertex at the source CC. \protect\poolball{3} the \textit{future} LCO is set, the runtime resumes the prior action state.}
    \label{fig:alloc-process}
  \end{minipage}
\end{figure*}

\begin{figure*}
  \centering
  \includegraphics[width=1\linewidth]{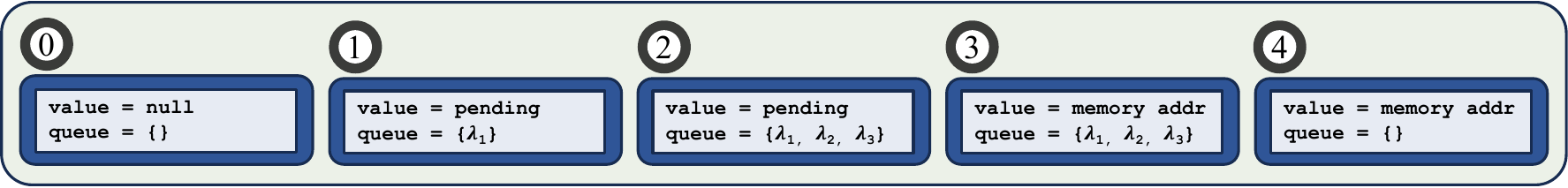}
  %\captionsetup{skip=0pt} % Reduce the space between figure and caption
  \caption{\texttt{ghost : (\textcolor{violet!80!black}{Future Pointer})}, a future of pointer type, as an exemplar shows the internal state of the future object as it is being set. \protect\poolball{0} null state. \protect\poolball{1} the first \langoperator{insert-edge-action} (see Listing \ref{lst:cca-insert-edge}) puts it in pending as it is being waiting to be set. \protect\poolball{2} some actions that have dependency on this future arrive and their related tasks are enqueued in the form of a closure task. \protect\poolball{3} a continuation from a remote compute cell, returned in the form of an action, sets the future with the address of the newly allocated remote memory space. \protect\poolball{4} depended tasks are scheduled, and the future queue is emptied.}
\label{fig:future-process}
\end{figure*}

\lstset{
    language=Racket,
    basicstyle=\ttfamily\scriptsize, % \ttfamily\footnotesize,
    % keywordstyle=\color{blue}\ttfamily,
    stringstyle=\color{red}\ttfamily,
    % commentstyle=\color{orange!80!black}\ttfamily,
    commentstyle=\color{green!40!black}\ttfamily,
    morecomment=[l][\color{magenta}]{\#},
    showstringspaces=false,
    breaklines=true,
    %backgroundcolor=\color{gray!5},
    numbers=left,
    numberstyle=\tiny\color{gray},
    numbersep=2pt,
    upquote=true,
    % frame=single,
}

\begin{lstlisting}[language=Racket,caption={Insert an edge and update the BFS level along the newly added edge.},label=lst:cca-insert-edge-simple,xleftmargin=0.3cm, escapeinside={*@}{@*}]
(define insert-edge-action ;; v: target vertex, e: edge
  (lambda ([v : (Pointer vertex)] [e : edge])
    (begin
      (insert-edge v e) ;; Insert the edge in edgelist.
      ;; Inform the dst vertex about this new edge
      ;; only if this src vertex has a valid BFS level.
      (if (not (= (vertex-level v) max-level)) 
        (propagate (bfs-action 
                    (list (edge-addr e) 
                          (+ (vertex-level v) 1))))))))
\end{lstlisting}

\begin{lstlisting}[language=Racket,caption=Breadth first search action.,label=lst:cca-bfs,xleftmargin=0.3cm]
;; Breadth First Search Action
(define bfs-action ;; v: target vertex, lvl: level in
  (lambda ([v : (Pointer vertex)] [lvl : Integer])
    (if (> (vertex-level v) lvl)
      (begin
        ;; Perform work.
        (set-vertex-level! v lvl) 
        ;; Diffusion occurs.
        (for-each 
          (lambda (e)
            ;; Get address of vertex along edge e.
            (let ([addr (edge-addr e)])
              ;; Send action along edge e.
              (propagate bfs-action (list addr (+ lvl 1)))))
          (vertex-edges v))
      ))))
\end{lstlisting}

\subsection{Continuation \& Future LCO}\label{subsec:continuation-future}

\begin{comment}
\begin{enumerate}
    \item When an edge is added and a subsequent action is called for dynamic graph processing.
    \item When a future holds memory allocation that is yet to be assigned.
    \item Our continuation doesn't really hold much context. The action finishes and a new action is invoked from there on. Perhaps some context for the future used in the ghost vertex malloc ???
\end{enumerate}

\textbf{Limitations:} 
\begin{enumerate}
    \item runtime context of action is not stored and therefore the user has to explicitly store any information locally before calling call/cc. ?? is still valid?
    \item recursive continuations are currently not supported due to parent action arguments being moved around along with the continuation. ????
    \item call/cc accepts an action and therefore the only time the arguments are moved around are if that actions is to be invoked on a remote compute cell. In the case of dynamic BFS the bfs-action will be invoked locally.
\end{enumerate}

\begin{enumerate}
    \item The continuation is created by the runtime using and LCO that is triggered when the continuation returns.
    \item The newly created LCO's memory addr is passed to the continuation for it to know where to return and set the LCO.
    \item The LCO contains a closure that contains runtime context of the previous action.
    \item The runtime context is created by the compiler in the form of lines of code following the call/cc and any data (variables) that must be needed for the new closure.
\end{enumerate}
\end{comment}

Listing \ref{lst:cca-insert-edge} demonstrates the use of continuation (\langkeyword{call/cc}) and \textit{future} LCO for the edge insertion. Continuation is used when a new ghost vertex must be allocated before the insertion can safely take place. Since the allocation operation will asynchronously happen on a remote compute cell, a continuation is created (Line \texttt{\#$16$-$18$}) that sets the ghost vertex with the address of the newly allocated memory. Internally the continuation is implemented by the compiler \footnote{In our implementation, we write this by hand, but eventually, it the compiler's job.} working together with the Runtime. The compiler first generates an anonymous \textit{action} that only includes lines of code (instructions) following \langkeyword{call/cc} keyword. It then injects code that asks the Runtime to \langoperator{propagate} the \langoperator{allocate} system \textit{action} with this anonymous \textit{action} as its return trigger. In this way the anonymous \textit{action} will be triggered when the continuation returns. The current \textit{action} immediately returns, meaning that lines \#$16$ and onwards are not executed. They will be executed when the continuation returns, triggering the anonymous \textit{action}. In the meantime, the Runtime schedules other tasks that may be present on this compute cell.

When the continuation returns with the address of the newly allocated ghost vertex, it triggers the anonymous \textit{action} that resumes the prior \textit{action} state that is waiting on this continuation. It sets the ghost vertex on line \#$16$. Figure \ref{fig:alloc-process} further graphically explains this process. Figure \ref{fig:future-process} demonstrates the internal details and functions of the \textit{future} LCO object, for a vertex, as it is being set.

\begin{lstlisting}[language=Racket,caption={Insert an edge and allocate a ghost vertex if needed. There can be two or more ghost vertices per RPVO to arbitrate, and the \langoperator{allocate} will require more arguments, these details are omitted for brevity.},label=lst:cca-insert-edge,xleftmargin=0.3cm, escapeinside={*@}{@*}]
(define insert-edge-action ;; v: target vertex, e: edge
  (lambda ([v : (Pointer vertex)] [e : edge])
    (if (vertex-has-room v) ;; If edgelist is not full.
      (begin
        (insert-edge v e) ;; Insert the edge in edgelist.
        ;; Inform the dst vertex about this new edge
        ;; only if this src vertex has a valid BFS level.
        (propagate (bfs-action ... ))
      )
      (begin ;; Else send the edge e to ghost vertex.
        ;; Check the ghost future has been fulfilled?
        (if (null? (vertex-ghost v)) 
          (begin ;; Ghost is not allocated yet.
            ;; Set future to pending.
            (future-pending! (vertex-ghost v))
            (set-future! (vertex-ghost v)
                  ;; Allocate memory using continuation.
                  (call/cc (allocate vertex)))) 
          ;; Else check whether the future is being 
          ;; fulfilled by a previous continuation.
          (if (future-pending? (vertex-ghost v))
            ;; Enqueue task in future
            (enqueue-future! (vertex-ghost v)
              (lambda 
                (propagate (insert-edge-action
                           (list (vertex-ghost v) e)))))
            ;; Else ghost exists, just recursively 
            ;; propagate the edge to ghost.
            (propagate (insert-edge-action 
                       (list (vertex-ghost v) e))))
        )))))
\end{lstlisting}

\section{Experimental Methodology}

We implement the ideas discussed in this paper using our simulator called the CCASimulator \cite{ccasimulator:online}. The simulator is high-level enough to be programmed using the \textit{diffusive programming model} and yet low-level enough to simulate individual message movements between CCs. In a single simulation cycle, a message can traverse one hop from one CC to a neighboring CC. We make this assumption since AM-CCA channel links are $256$ bit wide and can easily send the small \textit{messages} of our tested applications in a single flit cycle. Simultaneously, a single CC, can perform either of the two operations: 1) a computing instruction, which is contained in the \textit{action}, or 2) the creation and staging of a new message when an instance of \langkeyword{propagate} is called. The simulator employs turn-restricted routing that is deadlock free and always traverses the minimal path between source and destination \cite{TurnRestricted1992}. In particular, the YX dimension ordered routing that takes vertical paths first before turning horizontal. We verify the results for correctness against known results found using NetworkX \cite{NetworkX2008}. Assumption for the energy cost model remain the same as in our previous work \cite{chandio2024rhizomes}. Our asynchronous streaming dynamic BFS implementation is available at \cite{ccasimulator:online}.

\textbf{Datasets:} We perform our experiments using dynamic graphs from MIT's Streaming GraphChallenge \cite{StreamingChallenge2017}\cite{StreamingDataSets}. Table \ref{tab:graphdetails} provides details of the graph datasets used in our dynamic graph experiments. The graphs are constructed using two types of sampling methods: Edge and Snowball. In edge sampling, the edges are inserted as if they were formed or observed in the real world, while in Snowball sampling, the edges are inserted as they are discovered from a starting point \cite{graphChallenge2017}.

\begin{table*}
  \caption{Details of the GraphChallenge input dynamic graphs.}
  \label{tab:graphdetails}
  \centering  
  \begin{tabular}{|c|c|c|c|c|c|c|c|c|c|c|c|c|}
    \toprule
    \hline    
    & \textbf{Sampling} & \multicolumn{10}{|c|}{\textbf{Edges Per Streaming Increment}} & \textbf{Final} \\
    \cline{3-12}
    \textbf{Vertices} &\textbf{Type}  & \textbf{1} & \textbf{2} & \textbf{3} & \textbf{4} & \textbf{5} & \textbf{6} & \textbf{7} & \textbf{8} & \textbf{9} & \textbf{10} & \textbf{Edges}\\
    \hline\hline
    \numtothousand{50000} & Edge & \numtothousandNoK{101682} & \numtothousandNoK{102012}  & \numtothousandNoK{101772} & \numtothousandNoK{101916} & \numtothousandNoK{101634} & \numtothousandNoK{101254} & \numtothousandNoK{101809}  & \numtothousandNoK{102076} & \numtothousandNoK{101645} & \numtothousandNoK{102239} & \numtomillion{1018039} \\
    \hline
    \numtothousand{50000} & Snowball &  \numtothousandNoK{37315} & \numtothousandNoK{29238}  & \numtothousandNoK{47983} & \numtothousandNoK{68183} & \numtothousandNoK{87863} & \numtothousandNoK{108642} & \numtothousandNoK{129477}  & \numtothousandNoK{149413} & \numtothousandNoK{169416} & \numtothousandNoK{190509} & \numtomillion{1018039} \\
    \hline
    \numtothousand{500000} & Edge & \numtothousandNoK{1016373} & \numtothousandNoK{1016853}  & \numtothousandNoK{1015533} & \numtothousandNoK{1018007} & \numtothousandNoK{1018340} & \numtothousandNoK{1017923} & \numtothousandNoK{1016834}  & \numtothousandNoK{1019103} & \numtothousandNoK{1016846} & \numtothousandNoK{1018701} & \numtomillion{10174513} \\
    \hline
    \numtothousand{500000} & Snowball & \numtothousandNoK{222847} & \numtothousandNoK{328912}  & \numtothousandNoK{513890} & \numtothousandNoK{709723} & \numtothousandNoK{904420} & \numtothousandNoK{1101941} & \numtothousandNoK{1297078}  & \numtothousandNoK{1501559} & \numtothousandNoK{1698228} & \numtothousandNoK{1895915} & \numtomillion{10174513} \\
    \hline
    \bottomrule
    \multicolumn{13}{l}{\footnotesize There are ten increments to the graph each inserting a number of new edges. K is thousand, and M is million.}\\
  \end{tabular}
\end{table*}

\begin{table}
  \caption{Estimates of energy consumption and time taken for the $32\times32$ chip using $590 mm^2$ area and clocked at 1 \si{\giga\hertz}.}
  \label{tab:energy-time}
  \centering  
  \begin{tabular}{|c|c|c|c|c|c|}
    \toprule
    \hline    
    & \textbf{Sampling} & \multicolumn{2}{|c|}{\textbf{Ingestion}} & \multicolumn{2}{|c|}{\textbf{Ingestion \& BFS}} \\
    \cline{3-6}
    \textbf{Vertices} &\textbf{Type}  & \textbf{Energy} & \textbf{Time} & \textbf{Energy} & \textbf{Time} \\
    \hline\hline
    \numtothousand{50000} & Edge & 1355 \si{\micro\joule} & 22 \si{\micro\second} & 4669 \si{\micro\joule} & 68 \si{\micro\second} \\
    \hline
    \numtothousand{50000} & Snowball &  1357 \si{\micro\joule} & 25 \si{\micro\second} & 2929 \si{\micro\joule} & 43 \si{\micro\second} \\
    \hline
    \numtothousand{500000} & Edge & 13480 \si{\micro\joule} & 206 \si{\micro\second} & 50274 \si{\micro\joule} & 694 \si{\micro\second} \\
    \hline
    \numtothousand{500000} & Snowball & 13498 \si{\micro\joule} & 232 \si{\micro\second} & 32895 \si{\micro\joule} & 448 \si{\micro\second} \\
    \hline
    \bottomrule
    \multicolumn{6}{l}{\footnotesize \si{\micro\joule} is microjoule, and \si{\micro\second} is microseconds.}\\
  \end{tabular}
\end{table}

\begin{figure}
  \centering
  \begin{subfigure}{0.49\linewidth}
    \centering
    \includegraphics[width=\linewidth]{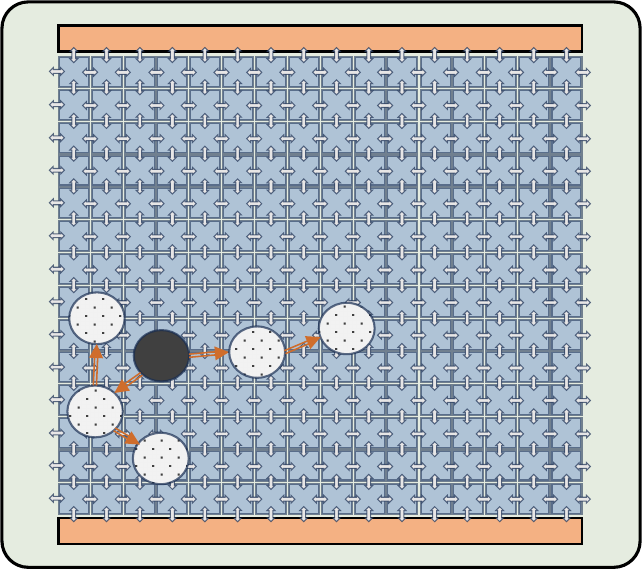}
    \caption{Vicinity Allocator.}
    \label{fig:RPVO-allocator}
  \end{subfigure}
   \begin{subfigure}{0.49\linewidth}
   \centering
   \includegraphics[width=\linewidth]{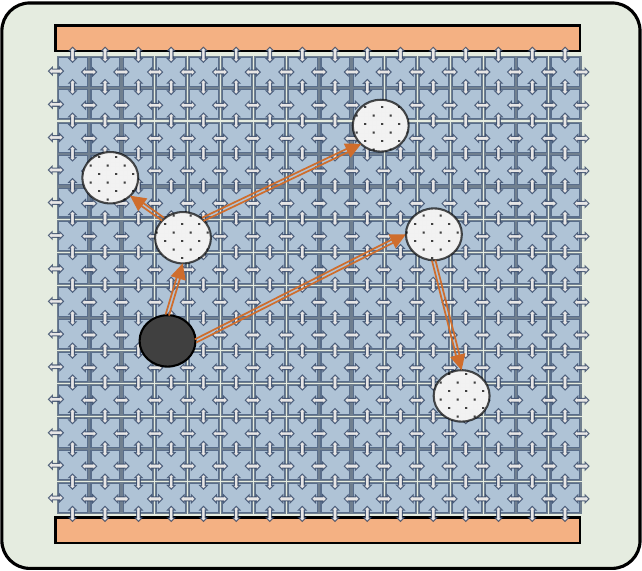}
   \caption{Random Allocator.}
   \label{fig:random-allocator}
  \end{subfigure}
  \caption{Vertex object allocation policy: (a) Localize ghost vertices in Compute Cells nearby, and (b) No regard to locality of ghost vertices.}
    \label{fig:allocators-mini}
\end{figure}

\textbf{Graph Construction:}
The graph is constructed by first allocating the root RPVO objects on the AM-CCA chip. Once the vertices are allocated and their addresses are known the edges are ingested into the chip by sending a message containing the edge using the \langoperator{insert-edge-action} of Listing \ref{lst:cca-insert-edge}. These ingestion messages originate from IO Cells in the IO channels, which read the edges and then distribute them among their respective IO Cells. When the computation starts, every cycle, each IO Cell reads an edge, creates the corresponding action registered with \texttt{INSERT\_ACTION}, and sends it to its connected Compute Cell (CC). For ghost vertex allocation, we keep the allocation nearby using the \textit{Vicinity Allocator} thus keeping the intra-vertex operation latency to a minimum. In particular, we set it to not be more than $2$ hops away from the originating CC. Figure \ref{fig:allocators-mini} conceptually shows this idea. It is contrasted with the \textit{Random Allocator} that randomly disperses the ghost vertices.

\begin{figure}
  \centering
  \begin{subfigure}{1\linewidth}
    \centering
    \includegraphics[width=\linewidth]{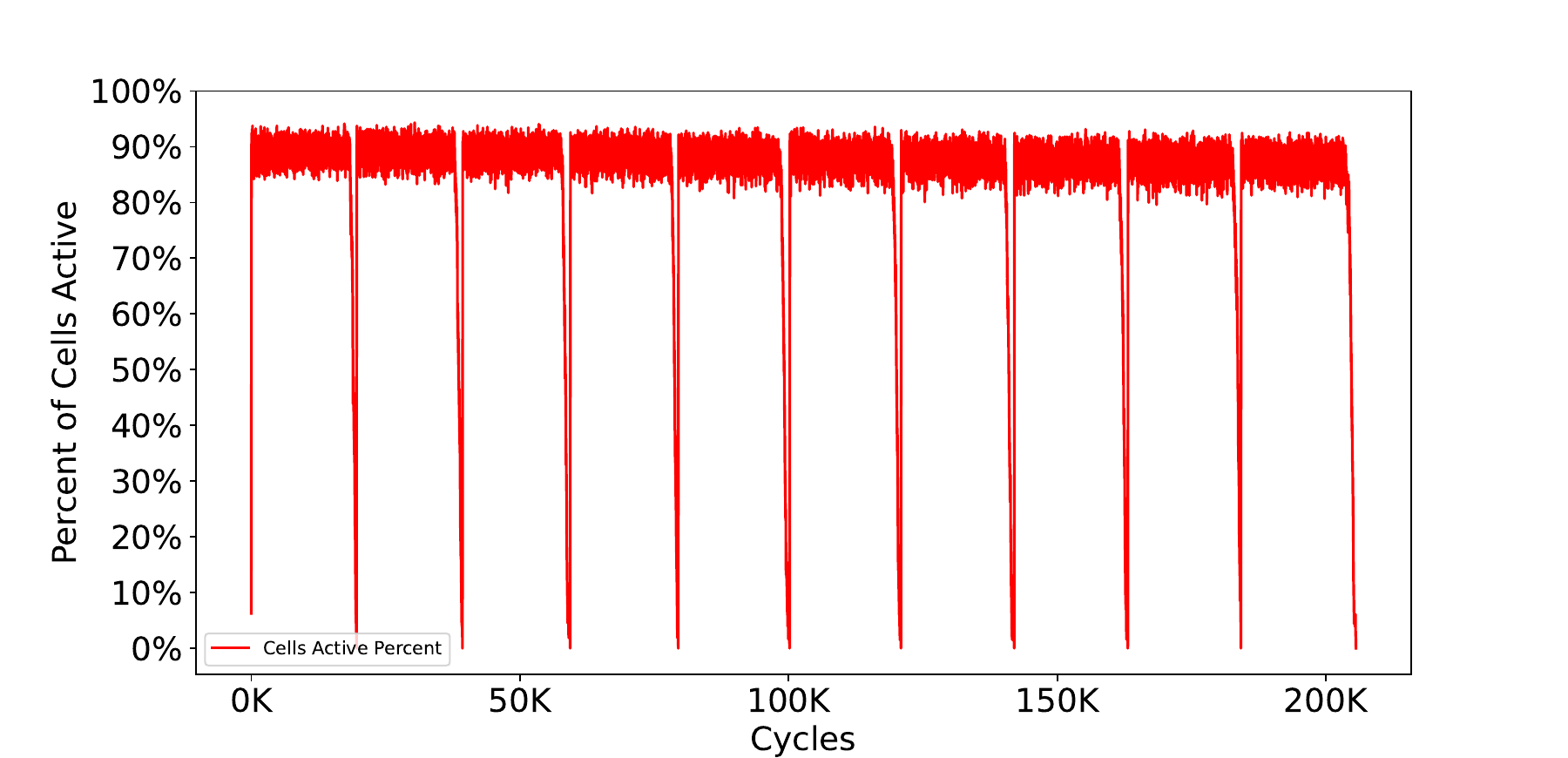}
    \caption{Edge Sampling.}
    \label{fig:streaming-edge-active-edge-500K}
  \end{subfigure}
  \hfill
  \begin{subfigure}{1\linewidth}
    \centering
    \includegraphics[width=\linewidth]{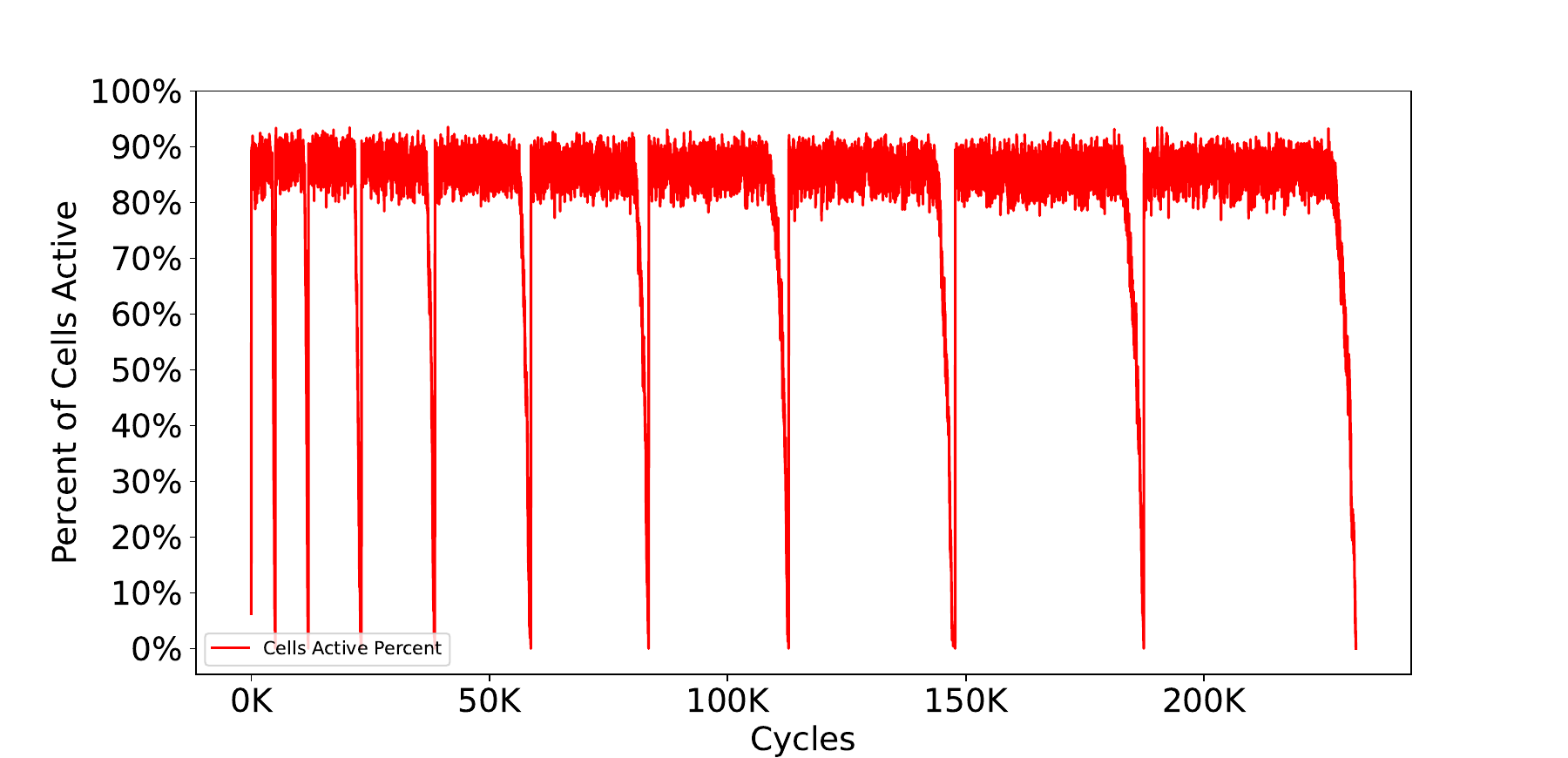}
    \caption{Snowball Sampling.}
    \label{fig:streaming-edge-active-snowball-500K}
  \end{subfigure}
  \caption{Streaming Edge Ingestion Only: activation status of compute cells per cycle of a $32 \times 32$ chip for graph with $500K$ vertices.}
  \label{fig:streaming-edge-active}
\end{figure}

\begin{figure}
  \centering
  \begin{subfigure}{1\linewidth}
    \centering
    \includegraphics[width=\linewidth]{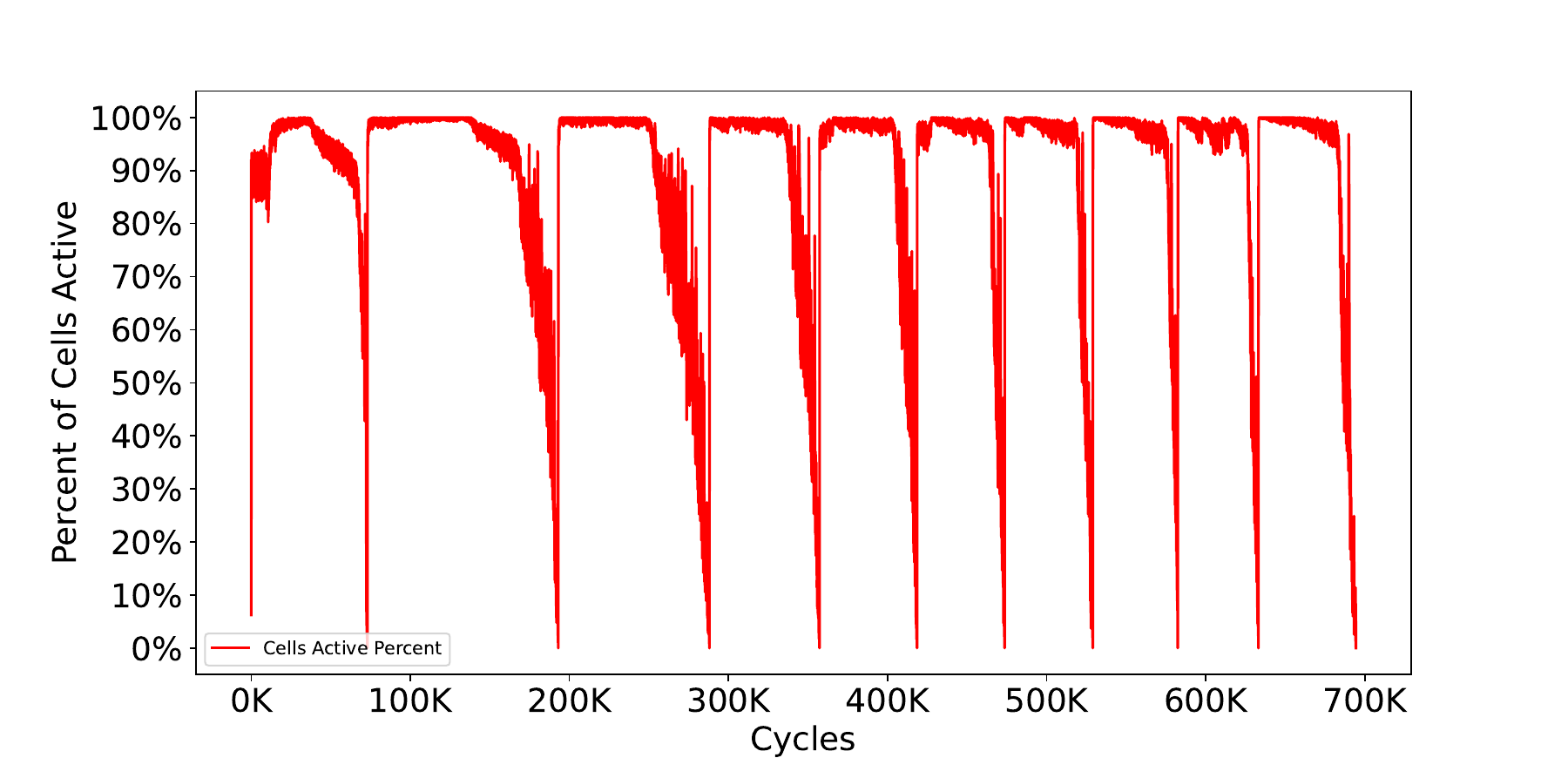}
    \caption{Edge Sampling.}
    \label{fig:streaming-bfs-active-edge-500K}
  \end{subfigure}
  \hfill
  \begin{subfigure}{1\linewidth}
    \centering
    \includegraphics[width=\linewidth]{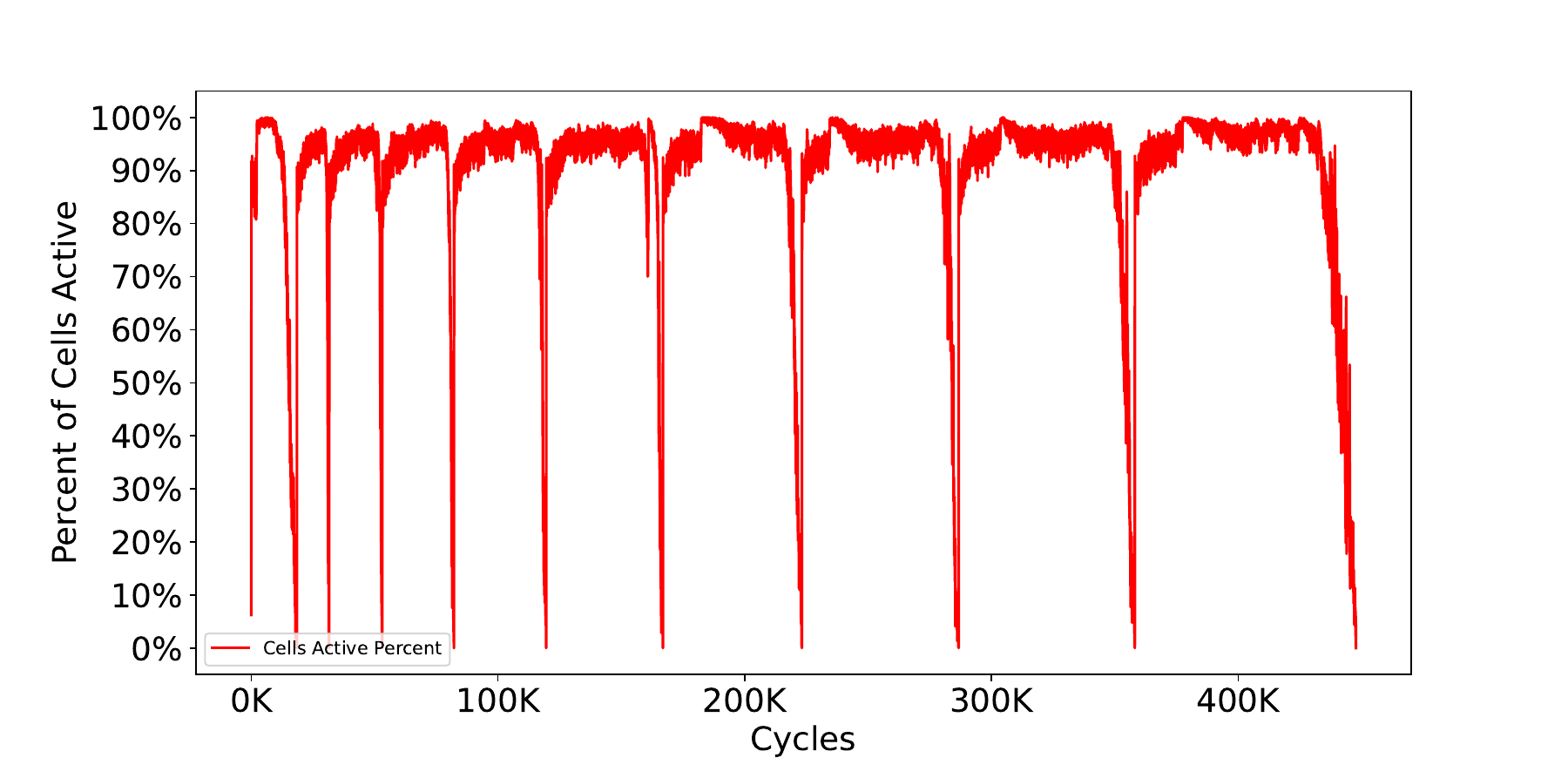}
    \caption{Snowball Sampling.}
    \label{fig:streaming-bfs-active-snowball-500K}
  \end{subfigure}
  \caption{Streaming Edge Ingestion with BFS: activation status of compute cells per cycle of a $32 \times 32$ chip for graph with $500K$ vertices.}
  \label{fig:streaming-bfs-active}
\end{figure}

\section{Results}
\begin{figure}
  \centering
  \begin{subfigure}{.8\linewidth}
    \centering
    \includegraphics[width=\linewidth]{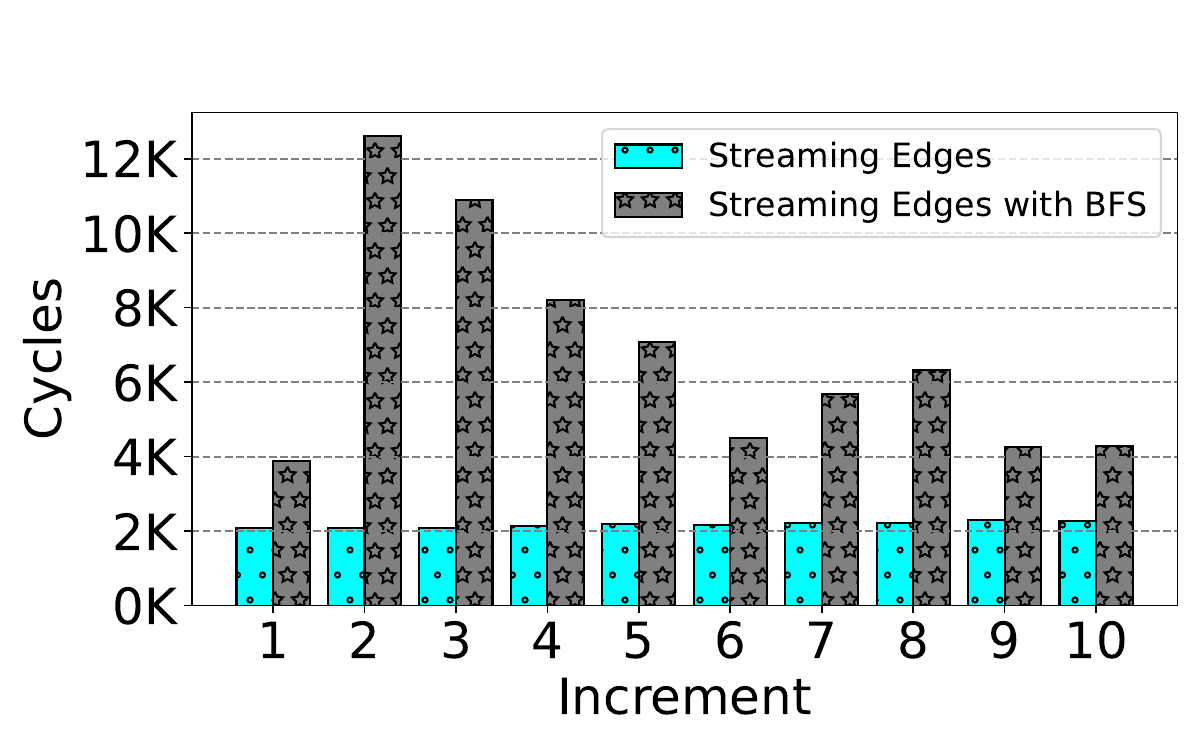}
    \caption{Edge Sampling.}
    \label{fig:edge-sample-50k}
  \end{subfigure}
  \hfill
  \begin{subfigure}{.8\linewidth}
    \centering
    \includegraphics[width=\linewidth]{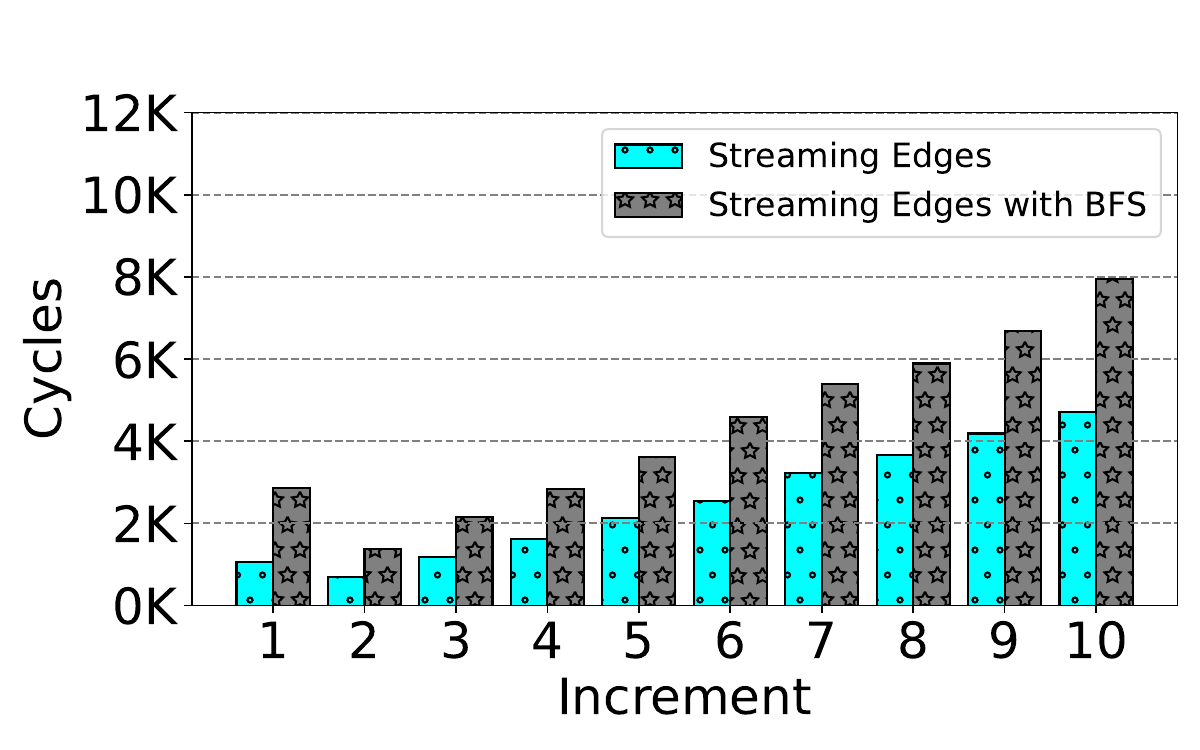}
    \caption{Snowball Sampling.}
    \label{fig:snowball-sample-50k}
  \end{subfigure}
  \caption{Time taken in simulation cycles on a $32 \times 32$ chip for graph with $50K$ vertices.}
  \label{fig:dynamic-bfs-performance-50K}
\end{figure}

\begin{figure}
  \centering
  \begin{subfigure}{.8\linewidth}
    \centering
    \includegraphics[width=\linewidth]{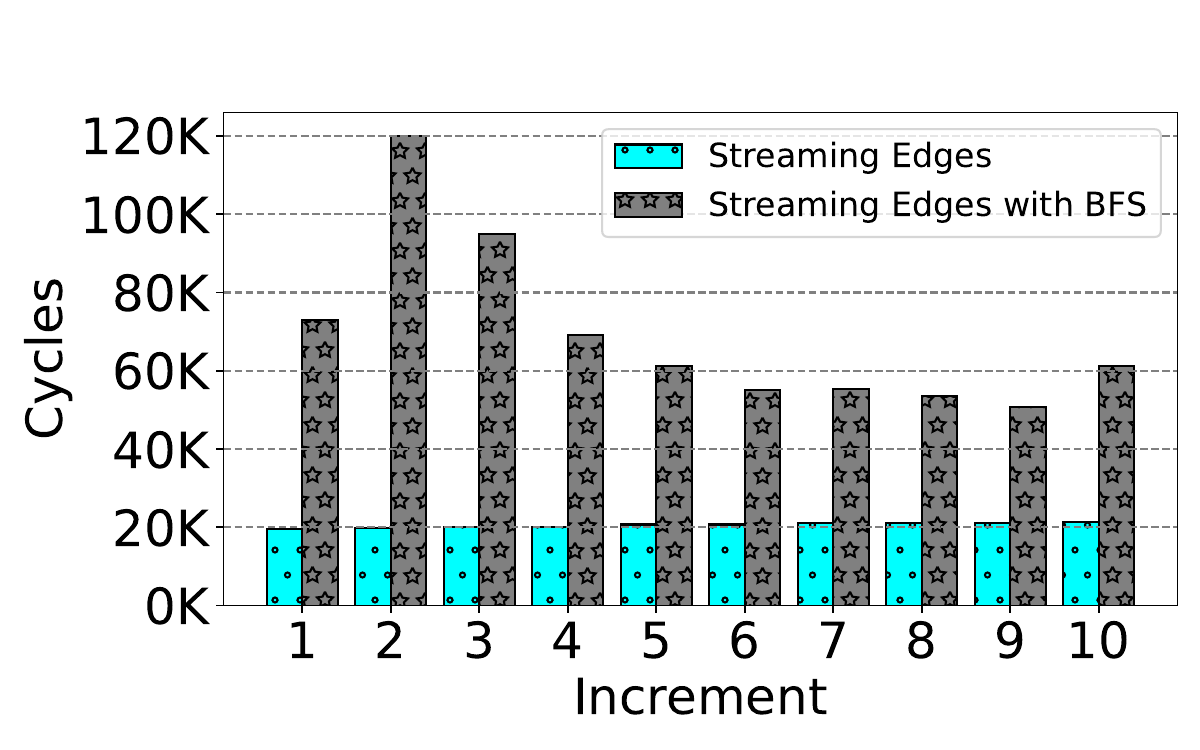}
    \caption{Edge Sampling.}
    \label{fig:edgesample-500K-32}
  \end{subfigure}
  \hfill
  \begin{subfigure}{.8\linewidth}
    \centering
    \includegraphics[width=\linewidth]{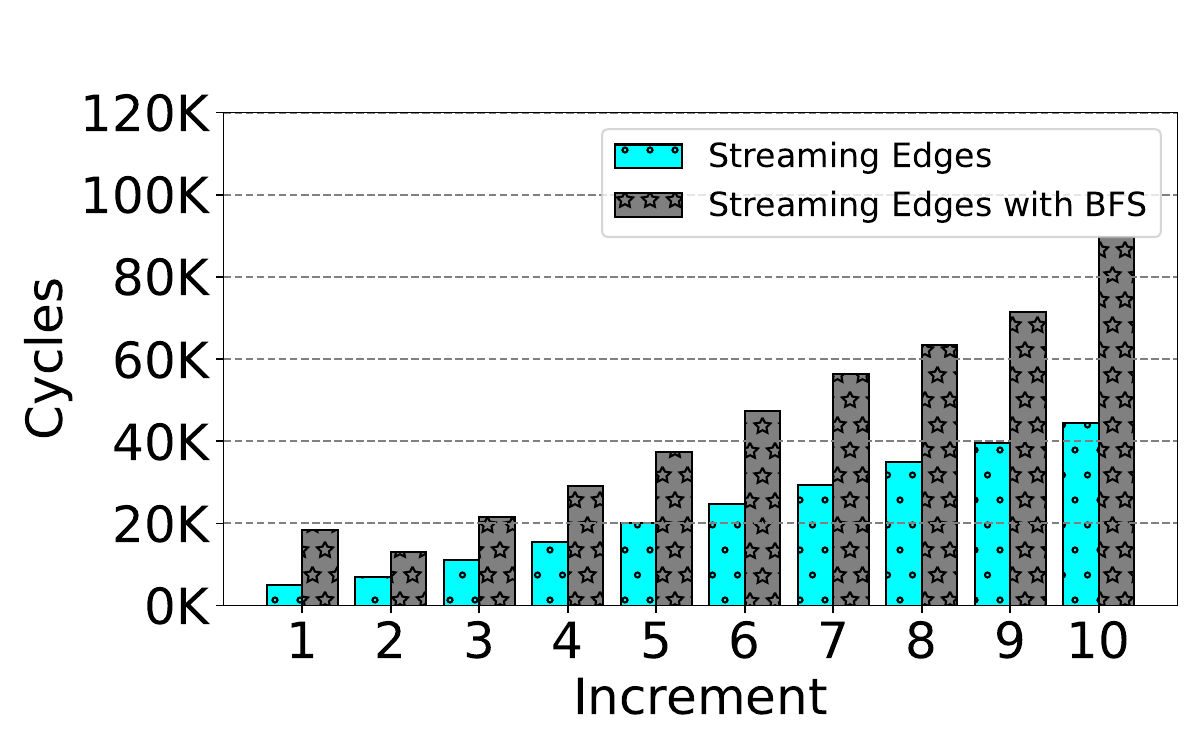}
    \caption{Snowball Sampling.}
    \label{fig:snowball-500K-32}
  \end{subfigure}
  \caption{Time taken in simulation cycles on a $32 \times 32$ chip for graph with $500K$ vertices.}
  \label{fig:dynamic-bfs-performance-500K}
\end{figure}

We run our asynchronous streaming dynamic BFS using the graphs of Table \ref{tab:graphdetails} on a $32\times32$ AM-CCA chip. Figure \ref{fig:dynamic-bfs-performance-50K} and Figure \ref{fig:dynamic-bfs-performance-500K} show the simulation cycles taken per dynamic graph increment for graph sizes $50K$ and $500K$, respectively. To differentiate between the time taken for data ingestion and BFS computation, we performed a separate experiment by disabling the subsequent propagation of \langoperator{bfs-action} when an edge is inserted. It provided the time taken, in simulation cycles, for only the streaming edge insertion per increment. Although, the complex interaction between \langoperator{insert-edge-action} \textit{actions} and \langoperator{bfs-action} \textit{actions} cannot be isolated completely, this approach provides a reference for estimating the additional time required to perform the BFS with the newly added edges and the previously computed state.

As expected, the ingestion time per increment for edge sampling remains similar, and for snowball sampling grows with each increment. It is due to the size of increment with edge sampling having similar amount of edges per increment and snowball sampling having increasing amount of edges. Although, the amount of edges for edge sampling remain same per increment, the execution time varies. This is due to the random sampling of edges that when inserted, randomly activate the vertices for BFS. This results in more \langoperator{bfs-action}s being created and propagated randomly varying the amount of work. Executions for snowball sampling don't show this behavior since snowball sampling itself was derived from a breath first traversal of the graph that addes edges with monotonically increasing BFS levels.

Table \ref{tab:energy-time} shows the estimated energy consumption in microjoules, and execution time in microseconds when the chip is clocked at $1$ \si{\giga\hertz}. For ingestion only, snowball sampling takes slightly longer due to the nature of edge insertion that targets a few vertices every increment. For a given increment, most edges are inserted from a few vertices, mostly in that frontier, which leads to congestion on a few compute cells that host these vertices.

Finally, we plot the behavior of the system in terms of the activation of compute cells per simulation cycle. Figure \ref{fig:streaming-edge-active} and Figure \ref{fig:streaming-bfs-active} show the chip active status for streaming data ingestion only and for streaming ingestion with BFS, respectively. We also create visual animations of the system from the trace of the simulation showing how streaming dynamic BFS transfers parallel control over the cellular grid of the AM-CCA chip. These animation are available at our repository \cite{ccasimulator:online}.

\section{Conclusion \& Future Work}
The paper presented structures and techniques geared towards co-designing asynchronous, decentralized dynamic graph processing for fine-grain memory-driven architectures. As the edges were streamed into the system they were converted into \textit{actions}, to pass data and control, and enable streaming dynamic updates to the graph structure. It resulted in very fine-grain updates to a hierarchical dynamic vertex data structure called RPVO, which subsequently triggered a user application action, BFS in particular, to update the results of any previous computation without recomputing from scratch.

Having build the scaffolding and demonstrated its capabilities, a natural future path is to design and implement more complex message-driven streaming dynamic algorithms. These include, but not limited to, Triangle Counting, Jaccard Coefficient, and Stochastic Block Partition.


\begin{comment}
\begin{acks}
\textbf{To be added:}
\end{acks}
\end{comment}
\bibliographystyle{ACM-Reference-Format}
\bibliography{Reference}

\end{document}